\documentclass[a4paper, 11pt]{llncs}
\usepackage{algorithm,algorithmic}
\usepackage{multicol}
\usepackage{epsfig}
\usepackage[latin2]{inputenc}
\usepackage{amsmath}
\usepackage{xspace}
\usepackage{amsfonts}

\usepackage[T1]{fontenc}

\newcommand{\SETOF}[2]{\{#1\;|\;#2\}}
\newcommand{\bit}{\{0,1\}}
\newcommand{\mspr}{{\sc Min-SPR}\xspace}

\newtheorem{fact}{Fact}

\begin{document}
\title{Shortest paths between shortest paths\\ and independent sets\thanks{A preliminary version of this work has been accepted to
IWOCA 2010, 21st International Workshop on Combinatorial Algorithms, London, 26-28 July 2010~\cite{IWOCA2010}.}}

\author{Marcin~Kami\'nski$^1$\thanks{Charg\'e de Recherches du FRS-FNRS.} \and Paul Medvedev$^2$ \and Martin Milani\v c$^3$\thanks{Supported in part by ``Agencija za raziskovalno dejavnost Republike Slovenije'', research program P1-0285.}}
\institute{ $^1$D\'epartement d'Informatique, Universit\'e Libre de Bruxelles, Brussels, Belgium \\
\texttt{Marcin.Kaminski@ulb.ac.be}\\
$^2$Department of Computer Science, University of Toronto, Toronto, Canada\\
\texttt{pashadag@cs.toronto.edu}\\
$^3$FAMNIT and PINT, University of Primorska, Koper, Slovenia\\
\texttt{martin.milanic@upr.si}}

%\date{}
\maketitle
    \begin{abstract}
    We study problems of reconfiguration of shortest paths %and independent sets
    in graphs. We prove that the shortest reconfiguration sequence can be exponential in the size of the graph and that it is NP-hard to compute the shortest reconfiguration sequence even when we know that the sequence has polynomial length. Moreover, we also study reconfiguration of independent sets in three different models and analyze relationships between these models, observing that shortest path reconfiguration is a special case of independent set reconfiguration in perfect graphs, under any of the three models. Finally, we give polynomial results for restricted classes of graphs (even-hole-free and $P_4$-free graphs).
 \end{abstract}

%\bigskip
\section{Introduction}\label{sec-intro}
One of the biggest impacts of algorithmic graph theory has been
its usefulness in modeling real-world problems, where
the domain of the problem is modeled as a graph and
the constraints on the solution define feasible solutions.
For example, consider the problem of routing a certain commodity between two nodes  in a transportation network,
using as few hops as possible.
The transportation network can be modeled as a graph, each route can be modeled as a path, and
the feasible solutions are all the shortest paths between the two nodes.
Traditionally, the real-world user first defines a problem instance and then uses an algorithm to find a feasible solution
which she then ``implements'' in the real world.
However, some real-world situations do not follow this simple paradigm and are more dynamic because
they allow the solution to ``evolve'' over time.
For example, consider the situation where the commodity is already being transferred along a
shortest route,
but the operator has been instructed to use a different route, which is also a shortest path.
She can physically switch the route only one node at a time, but does not wish to interrupt the transfer.
Thus, she would like to switch between the two routes in as few steps as possible, while maintaining a
shortest path route at every intermediate step.

In general, this type of situation gives rise to a {\em reconfiguration} framework, where
we consider an algorithmic problem $\cal P$ and a way of transforming one feasible solution of an instance $I$ of $\cal P$ to another
(\emph{reconfiguration rule}).
Given two feasible solutions $s_1, s_k$ of $I$,
we want to find a \emph{reconfiguration sequence} $s_1, \ldots, s_k$
such that each $s_i$ ($1\leq i \leq k$) is a feasible solution of $I$, and
the transition between $s_i$ and $s_{i+1}$ is allowed by the reconfiguration rule.
An alternate definition is via the \emph{reconfiguration graph}, where
the vertices are the feasible solutions of $I$, and
two solutions are adjacent if and only if one can be obtained from the other by the reconfiguration rule.
The reconfiguration sequence is then a path between $s_1$ and $s_k$ in the reconfiguration graph.
We can then ask for the shortest reconfiguration sequence, or, in the {\em reconfigurability problem},
to simply check if the two solutions are {\em reconfigurable} (i.e., if such a sequence exists).

The reconfiguration framework has recently been applied in a number of settings,
including vertex coloring \cite{CerecedaHJ08,CerecedaHJ09,BonsmaCHJ07,BonsmaC09},
list-edge coloring \cite{ItoKD09},
clique, set cover, integer programming,
matching, spanning tree, matroid bases~\cite{ItoDHPSUU08},
block puzzles \cite{HearnD05}, independent set \cite{HearnD05,ItoDHPSUU08},
and satisfiability \cite{GopalanKMP09}.
In the well-studied vertex coloring problem, for example,
we are given two $k$-colorings of a graph, and the reconfiguration rule allows to change the color of a single vertex.
In a different example, we are given two independent sets, which we imagine to be two sets of tokens placed on the vertices,
and the reconfiguration rule is to slide a single token along an edge ({\em token sliding}).

Though the complexities of each of the many reconfiguration problems may each be studied independently,
a fundamental question is whether there exists any systematic relationship between
the complexity of the original problem and that of its reconfigurability problem.
To this end, current studies have revealed a pattern where most ``natural'' problems in P have their reconfigurability problems in P as well, while
problems whose reconfigurability versions are at least NP-hard are NP-complete.
For example, spanning tree, matching, and matroid problems in general (all in P) lead to polynomially solvable
reconfigurability problems, while the reconfigurability of independent set,
set cover, and integer programming (all NP-complete) are PSPACE-complete \cite{ItoDHPSUU08}.
Another example is satisfiability, where Gopalan et al. \cite{GopalanKMP09}
showed that reconfigurability instances arising from tight relations---a class for which it is easy to determine if the formula is satisfiable---can be solved in linear time; on the other hand, reconfigurability is PSPACE-complete for the class of formulas arising
from non-tight relations.

Ito et al. \cite{ItoDHPSUU08} have conjectured that this relationship is not true in general, and that there exist problems in P which give rise, in a natural way, to NP-hard reconfigurability problems. Indeed, the problem of deciding whether two $k$-colorings are reconfigurable is PSPACE-complete for (i) bipartite graphs and $k \geq 4$, and (ii) planar graphs, for $4 \leq k \leq 6$ \cite{BonsmaC09}. Clearly, 4-coloring of bipartite or planar graphs is in P. However, these are not ``natural'' problems in the sense that the colorings are not optimal. It is interesting to ask if there exists a ``natural'' problem in P whose reconfiguration version is NP-hard.

Another systematic relationship that has been pursued is between the complexity of a reconfigurability problem
and the diameter of the reconfiguration graph.
When the diameter is polynomial, a reconfiguration sequence is a trivial certificate for the reconfigurability of two instances, guaranteeing that the
problem is in NP. %; when it is exponential, this naive certificate does not work.
However, current evidence further suggests that for reconfigurability problems that are solvable in polynomial time,
the diameter is also polynomial. %; for those that are PSPACE-hard, it can be exponential.
In the study of $k$-coloring, it was found  that for $k \leq 3$,
the reconfigurability problem is solvable in polynomial time and the diameter of the reconfiguration graph
is at most quadratic in the number of vertices of the colored graph.
%; for $k>3$, the reconfigurability problem is PSPACE-complete and that there exist instances whose diameter is exponentially large.
For satisfiability, the formulas built from tight relations (whose reconfigurability is polynomial)
lead to reconfiguration graphs with linear diameter
%, while there are examples of non-tight relations for which it is exponential
\cite{GopalanKMP09}.
We are not aware of any natural problems with the property that the diameter can be exponential
while reconfigurability can be decided in polynomial time\footnote{For a very artificial one,
consider the problem in which instances are $n$-bit words and two instances are adjacent if they differ by 1 modulo $2^n$.
The diameter of the reconfiguration graph is $2^{n-1}$ but all pairs of instances are reconfigurable.};
however, such an example, if found, would indicate that the diameter cannot serve as a reliable indicator of the reconfigurability complexity.

In this paper, we introduce the reconfiguration version of the shortest path problem (Section~\ref{sec:hard}),
which arises naturally, such as in the routing example above.
We show that the reconfiguration graph can have exponential diameter, implying that the shortest path
reconfiguration problem probably breaks one of the two established patterns described above.
On the one hand, if reconfigurability of shortest paths can be decided in polynomial time,
then it is the first example of a reconfigurability problem in P with exponential diameter.
On the other hand, if it is NP-hard, it is the first example of a ``natural'' problem in P
whose reconfigurability version is NP-hard. %, proving a strengthening of the conjecture of \cite{ItoDHPSUU08}.
For these reasons, we believe that shortest path reconfiguration is an important problem to study, not only for its
practical application but also for our understanding of the systematic relationship
between the hardness of a problem, the diameter of its reconfiguration graph, and the hardness of its reconfigurability
problem.
Towards this end, we give a non-trivial reduction from SAT to show that
it is NP-hard to find the shortest reconfiguration sequence between two shortest paths
(however, the complexity status of the reconfigurability problem remains open).

We also study reconfiguration of independent sets, where, unlike many other problems, there is more than one natural reconfiguration rule.
In particular applications, for example, a threshold is specified that bounds the cardinality of the intermediate feasible solutions.
Based on this idea, Ito et al. \cite{ItoDHPSUU08} considered an alternative to token sliding called {\em token addition and removal}, where one is allowed
to either add or remove a token as long as there are at least $k-1$ tokens at any given time, for some $k$.
In this paper, we introduce {\em token jumping}, where one is allowed to move a single token to any other vertex.
The token jumping reconfiguration graph is often easier to analyze than the token addition and removal one,
since the cardinalities of two adjacent token sets are always the same.
However, we show that the two models are polynomially equivalent, allowing for an easier way to analyze
token addition and removal reconfiguration graphs
(Section~\ref{sec:rel}).

Finally, we show that reconfiguration of independent sets is a generalization of the reconfiguration of shortest paths;
our hardness result for shortest paths then implies that it is NP-hard to find the shortest reconfiguration sequence between two independent sets, even in perfect graphs (Section~\ref{sec:paths-indep-sets}).
We also identify two restricted graph classes where reconfigurability is easy --
even-hole-free graphs under token jumping, for which the reconfiguration graph is always connected, and
$P_4$-free graphs under token sliding (Section~\ref{sec:positive}). 

\section{Shortest path reconfiguration}\label{sec:hard}
%Let $G=(V,E)$ be a graph with two designated vertices $s$ and $t$.
 %and let $p$ be a shortest path between $s$ and $t$.
We define the reconfiguration rule for shortest paths in the natural way:
two shortest $(s,t)$-paths are adjacent in the reconfiguration graph of shortest $(s,t)$-paths if and only if
they differ, as sequences, in exactly one vertex.
%The related existence problem, \spr, is to simply determine whether $p_b$ and $p_e$ are reconfigurable.

\subsection{Instances with exponential diameter}
\begin{figure}
\centering\includegraphics[width=0.7\textwidth]{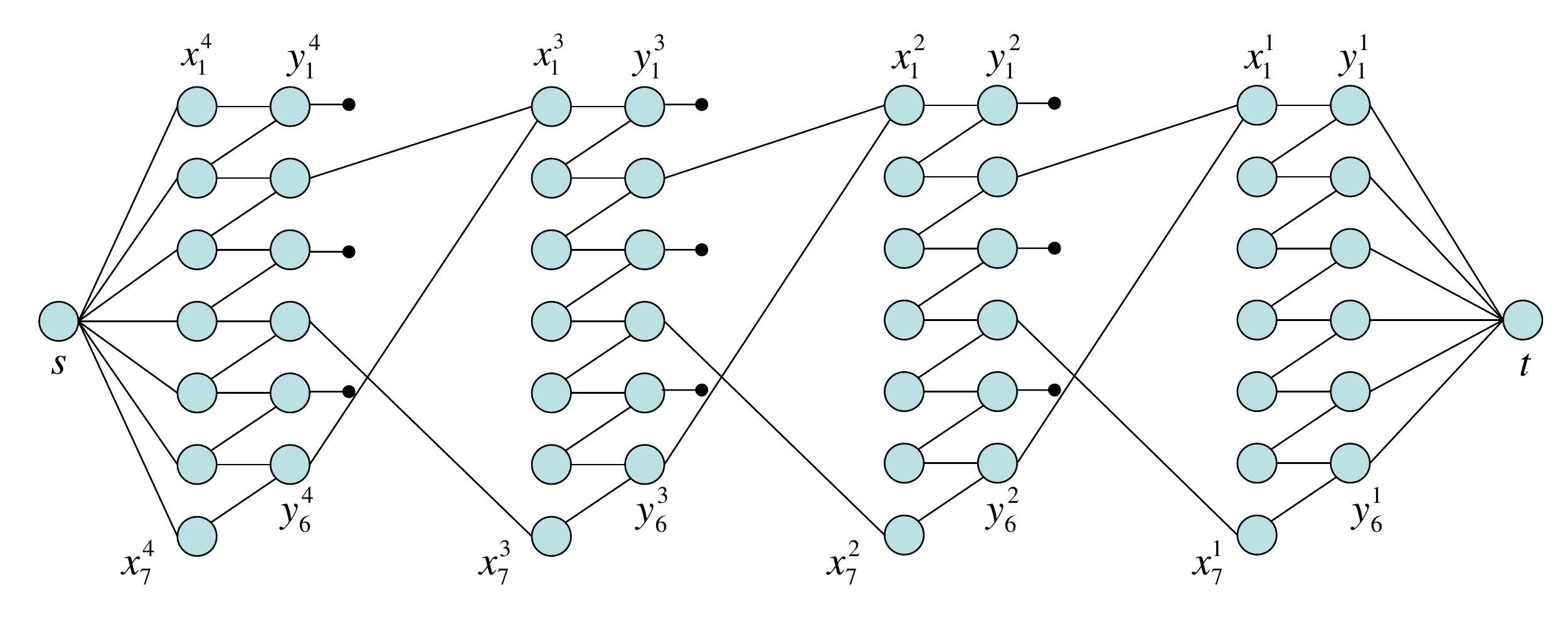}
%\vspace{-.5cm}
\caption{\label{fig:expEx} The graph $G^k$ for $k=4$, where the reconfiguration distance between
$p^k_b = s, x^k_1, y^k_1, \ldots, x^1_1, y^1_1, t$ and
$p^k_e = s, x^k_7, y^k_6, x^{k-1}_1, x^{k-1}_1, \ldots, x^1_1, y^1_1, t$ is $\Theta(2^k)$.
%the given two shortest paths is $\Theta(2^{k})$.
An edge with a circle end means that the vertex is connected to all the vertices in the next layer.
}
\end{figure}

\begin{sloppypar}
We now present a family of graphs $G^k$ whose size is linear in $k$ but the diameter of the reconfiguration graph is
$\Omega(2^k)$.
The graph $G^1$ contains vertices $\SETOF{x^1_i}{1\leq i \leq 7} \cup \SETOF{y^1_i}{1\leq i \leq 6} \cup \{s,t\}$ and edges
$\SETOF{(x^1_i, y^1_i), (x^1_{i+1}, y^1_i), (y^1_i,t) }{i \leq 6} \cup \SETOF{(s, x^1_i)}{1\leq i \leq 7}$.
The graph $G^k$ is defined recursively with vertices
$\SETOF{x^k_i}{1\leq i \leq 7} \cup \SETOF{y^k_i}{1\leq i \leq 6} \cup V(G^{k-1})$ and
the edges $\SETOF{(x^k_i, y^k_i), (x^k_{i+1}, y^k_i)}{i \leq 6 }
\cup \SETOF{(y^k_i, x^{k-1}_j)}{i\in\{1,3,5\}, j \leq 7}
\cup \{ (y^k_2, x^{k-1}_1), (y^k_4, x^{k-1}_7), (y^k_6, x^{k-1}_1) \}
\cup E(G^{k-1} \setminus \{s\})
\cup \SETOF{(s, x^k_i)}{1\leq i \leq 7}$
(see Figure~\ref{fig:expEx}).
Let $p^k_b = s, x^k_1, y^k_1, \ldots, x^1_1, y^1_1, t$, and let
$p^k_e = s, x^k_7, y^k_6, x^{k-1}_1, x^{k-1}_1, \ldots, x^1_1, y^1_1, t$.
We will consider the problem of reconfiguring $p^k_b$ to $p^k_e$ in $G^k$.
\end{sloppypar}
	
\begin{lemma}
	Let $p$ be a shortest path in $G^k$ that goes through $y^k_1$, and let $q$ be a path that goes through $y^k_6$.
	Then the reconfiguration distance between $p$ and $q$ is at least $9(2^{k} - 1) $.
\end{lemma}
\begin{proof}
	We prove by induction on $k$, where the base case is clear.
	Let $\rho = p_1, \ldots, p_n$ be the shortest reconfiguration sequence between $p$ and $q$.
	First, let $i'$ be the smallest integer such that $p_{i'+1}$ contains $y^k_4$,
	and let $i\leq i'$ be the smallest integer such that every path $p_i,\ldots,p_{i'}$ contains $y^k_3$.
	By construction, we know that $p_{i-1}$, and hence $p_i$, contains $y^{k-1}_1$ and $p_{i'+1}$, and hence $p_{i'}$,
	contains $y^{k-1}_6$.
	Hence, by the induction hypothesis, the length of this first phase, $i' - i + 1$, is at least $9(2^{k-1} -1)$.

	Next, let $j'$ be the smallest integer such that $p_{j'+1}$ contains $y^k_6$,
	and let $j\leq j'$ be the smallest integer such that every path $p_j,\ldots,p_{j'}$ contains $y^k_5$.
	By construction, we know that $p_{j-1}$, and hence $p_j$, contains $y^{k-1}_6$ and $p_{j'+1}$, and hence $p_{j'}$,
	contains $y^{k-1}_1$.
	Hence, by the induction hypothesis, the length of this second phase, $j' - j + 1$, is at least $9(2^{k-1} -1 )$.

	Observe from the graph construction that $\rho$ must always visit $y^k_{x-1}$ before visiting $y^k_x$, hence
	$i' < j$, and so the length of $\rho$ is at least the sum of the two phases plus the moves of the first
	and second vertex necessary to percolate $y^k_1$ down to $y^k_6$, proving the lemma.
	\qed
\end{proof}

On the other hand, there exists an asymptotically matching lower bound:
\begin{lemma}\label{lem:upper_bound}
	The reconfiguration distance between $p^k_b$ and $p^k_e$ is at most \hbox{$11(2^k - 1)$}.
\end{lemma}
\begin{proof}
	\begin{sloppypar}
	We prove by induction on $k$, where the base case is clear.
	It will be helpful to formally treat a reconfiguration sequence not as a sequence of paths but
	as a sequence of vertices, each one representing the switched vertex at that step.
	Applying the induction hypothesis, let $\rho$ be the shortest reconfiguration sequence in $G^{k-1}$, and let
	$rev(\rho)$ be that sequence in the reverse direction (from $p^{k-1}_e$ to $p^{k-1}_b$).
	We construct the sequence as $\rho'=x^k_2, y^k_2, x^k_3, y^k_3, \rho, x^k_4, y^k_4, x^k_5, y^k_5, rev(\rho), x^k_6, y^k_6, x^k_7$.
	This sequence of moves reconfigures $p^k_b$ into $p^k_e$ with the number of steps satisfying the lemma.
	\qed
	\end{sloppypar}
\end{proof}
We therefore have the following theorem:
\begin{theorem}\label{thm:expEx}
	The reconfiguration distance in $G^k$ between $p^k_b$ and $p^k_e$ is $\Theta(2^k)$.
\end{theorem}

\subsection{NP-Hardness of \mspr}
\begin{sloppypar}
Given $(G,s,t,p_b,p_e,k)$, where  $p_b$ and $p_e$ are shortest $(s,t)$-paths and $k$ is an integer,
the \mspr problem is to determine whether there is a reconfiguration sequence between $p_b$ and $p_e$ of length at most $k$.
Let $\phi$ be a formula with variables $x_1,\ldots, x_n$ and clauses $C_1,\ldots, C_m$.
We will create an instance $(G_\phi,s,t,p_b,p_e,2m(n+2))$ and show that $\phi$ is satisfiable if and only
this instance is in \mspr.
For ease of presentation, the graph $G_\phi$ will be directed.
However, our result holds for undirected graphs because the directed shortest $(s,t)$-paths in $G_\phi$
are exactly the shortest paths in the undirected version of $G_\phi$.
\end{sloppypar}

\begin{figure}
\centering\includegraphics[width=1.0\textwidth]{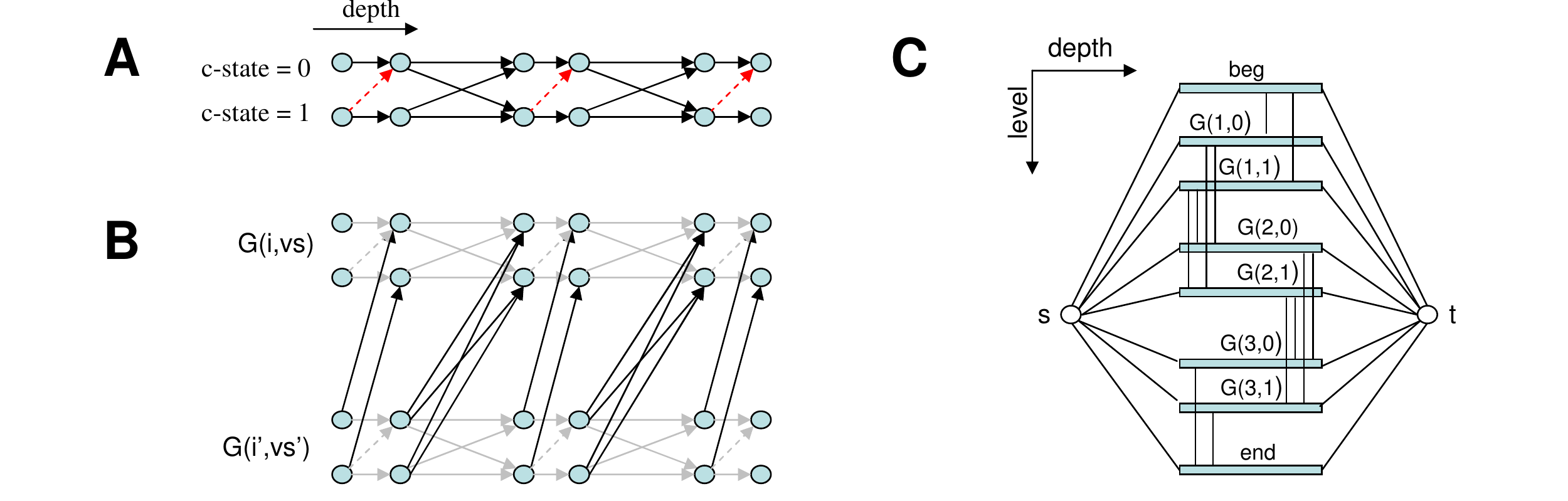}
\vspace{-0.5cm}
\caption{\label{fig:reduction}
The reduction from a formula $\phi$ to a graph $G_\phi$ for the case of three clauses and three variables.
Panel A shows the internal connections of a gadget, with
the potential formula edges that depend on $\phi$ given in red (dashed).
Panel B shows the way we connect two given gadgets, while
C shows the structure of the whole graph.
Each of the rectangles represents a gadget, with the lines showing which parts are connected together.
}
\end{figure}

For every variable $x_i$ and its possible value $vs \in \bit$, we build a gadget $G(i,vs)$.
The vertex set is $\SETOF{v(i,vs,cs,j)}{cs\in\bit, 1\leq j \leq 2m}$.
The values $i$, $vs$, $cs$, and $j$ for a vertex are referred to as its \emph{level}, \emph{v-state}, \emph{c-state}, and \emph{depth},
and denoted by $l(v)$, $vs(v)$, $cs(v)$, and $d(v)$, respectively.
For every $1\leq j \leq 2m - 1$, and every $cs$, there is an edge from
$v(i,vs,cs,j)$ to $v(i,vs,cs,j+1)$.
For all $1\leq j \leq m-1 $, there is an edge from $v(i,vs,0,2j)$ to $v(i,vs,1,2j+1)$,
and from $v(i,vs,1,2j)$ to $v(i,vs,0,2j+1)$.
We also add edges, called formula edges, that are formula dependent.
For all $j$, if $x_i = vs$ satisfies $C_j$, we add an edge from $v(i,vs,1,2j-1)$ to $v(i,vs,0,2j)$.
This gadget is shown in Figure~\ref{fig:reduction}A.

We now connect some of these gadgets together.
The gadgets we connect are $G(i,vs)$ to $G(i+1,0)$ and to $G(i+1,1)$, for all $i \leq n - 1$ and all $vs$.
Given two gadgets, $G(i,vs)$ and $G(i',vs')$, the meaning of connecting $G(i,vs)$ to $G(i',vs')$
is given as follows (shown in Figure \ref{fig:reduction}BC).
For all $j \leq 2m - 1$ and $cs$, there is an edge from $v(i',vs',cs,j-1)$ to $v(i,vs,cs,j)$.
Also, for all $j \leq m - 1$ and $cs$, there is an edge from $v(i',vs',cs,2j)$ to $v(i,vs,1 - cs,2j + 1)$.

We next add a begin and end gadget to the graph, consisting of vertices $beg_j$ and $end_j$, respectively, for $1\leq j \leq 2m$.
These are connected in a path, with edges $(beg_j, beg_{j+1})$ and $(end_j, end_{j+1})$ for $j\leq 2m -1$.
The level of the vertices in the begin (end) gadget is 0 ($i + 1$), the c-state is 0 (1), and the depth of $beg_{j}$ or $end_{j}$ is $j$.
For all $vs$, $j \leq 2m - 1$, there is an edge from $v(1,vs,0,j)$ to $beg_{j+1}$, and from $end_{j}$ to $v(n,vs,1,j+1)$.

Finally, we add a $s$ and $t$ vertex to the graph, and make an edge from $s$ to every depth 1 vertex, and from every depth $2m$ vertex to $t$.
The depth of $s$($t$) is defined to be 0 ($2m+1$).
We call the resulting directed graph $G_\phi$.
Let $p_b = s, beg_1, \ldots, beg_{2m}, t$ and $p_e= s, end_1, \ldots, end_{2m}, t$ be two paths in this graph.
Then, $(G_\phi, s, t, p_b, p_e, 2m(n+2))$ is the instance of the \mspr problem that we will consider here.

The intuition behind the reduction is that in order for the path to percolate down from $p_b$ to $p_e$
in a minimal number of steps, it must pass consecutively through exactly one of
$G(i,0)$ or $G(i,1)$ for every variable $x_i$.
The choice of which one corresponds to assigning $x_i$ the corresponding value.
Furthermore, each shortest path that goes through a gadget can visit the vertex at depth $2j$ with a c-state of 0 or 1.
This corresponds to having the $j^{\text{th}}$ clause satisfied or not.
Initially, the path goes only through vertices with c-state 0, and the only way to switch the c-state at a given depth
is via a formula edge.
By going through a gadget $G(i,vs)$, there is an opportunity to use the formula edges to switch the c-state of all
clauses that $x_i=vs$ would satisfy.
In order to reach the final path $p_e$, the c-state of all the vertices must be 1, hence all the clauses must be satisfied.

First, we will show that the reduction is sound.
Each edge $(a,b)$ is considered to be either \emph{odd} or \emph{even}, depending on the parity of $d(a)$.
We call edges that connect vertices on the same level (exactly those that belong to the same gadget) as \emph{intra-level},
while the edges that connect vertices on different levels are called \emph{inter-level}.
We say that a reconfiguration sequence \emph{visits} a vertex if there exists $p\in\rho$ that contains that vertex.

\begin{fact}\label{fct:sp}
	Let $e=(a,b)$ be an edge in $G_\phi$.
	The following follows directly from construction:
	\begin{enumerate}
		\item \label{fct1}$l(b) \leq l(a) \leq l(b) + 1$.
		\item \label{fct2}If $e$ is an intra-level odd edge, $cs(a)=0$ implies that $cs(b) = 0$.
		\item \label{fct3}If $e$ is an inter-level odd edge, then $cs(a) = cs(b)$.
		\item \label{fct4}If $e$ is a non-formula odd edge, then $cs(a) = cs(b)$.
		\item \label{fct5}If $e$ is intra-level, then $vs(a) = vs(b)$.
	\end{enumerate}
\end{fact}

These facts about $G_\phi$ capture most of the properties of the reduction that are needed to prove completeness and soundness.
We will first need some definitions.
Let $p = s, v_1, \ldots, v_{2m}, t$ be a shortest path and consider an arbitrary move that switches $v_i$ with $v'_i$.
The \emph{move graph} is the subgraph induced by $v_{i-1}, v_i, v'_i, v_{i+1}$,
referred to by the tuple $(v_{i-1}, v_i, v'_i, v_{i+1})$.
There are two kinds of moves, \emph{odd} and \emph{even}, depending on the parity of $d(v_i)$.

\begin{lemma}\label{lem:sp1}
	The length of a reconfiguration sequence is at least $2m(n+2)$.
	Moreover, each move in an sequence that has this length
	must either increase the c-state or the level of the switched vertex, but not both.
\end{lemma}
\begin{proof}
	Let $\Phi(p) = \sum_{v\in p \setminus \{s, t\} } l(v) + cs(v)$ be a potential function counting the sum levels and c-states for vertices in a path.
	Consider an arbitrary move $m$ and its move graph $(a,b,c,d)$.
	We will first show that $\Delta\Phi(m) = l(c) - l(b) + cs(c) - cs(b) \leq 1$.
	First consider the case that the move is odd.
	Applying Fact~\ref{fct:sp}.\ref{fct1} to the edges $(b,d)$ and $(c,d)$, we get that $l(c) \leq l(b) + 1$.
	Moreover, the only way to have $l(c) = l(b) + 1$ is for $(b,d)$ to be intra-level and $(c,d)$ to be inter-level.
	In this case, and further supposing $cs(b)=0$,
	Fact~\ref{fct:sp}.\ref{fct2} implies that $cs(d) = 0$, and Fact~\ref{fct:sp}.\ref{fct3} implies that $cs(c) = 0$.
	These facts imply $\Delta\Phi(m) \leq 1$.
	A similar argument, applied to the edges $(a,b)$ and $(a,c)$, holds for the case $m$ is even.
	Combined with the fact that $\Phi(p_b) = 0$ and $\Phi(p_e) = 2m(n+2)$, the lemma follows.
	\qed
\end{proof}

\begin{lemma}\label{lem:sppath}
	No path can contain two vertices with the same level but different v-state.
\end{lemma}
\begin{proof}
	In any path, the level of the vertices is non-increasing (by Fact~\ref{fct:sp}.\ref{fct1}).
	Therefore, all the vertices that have the same level must appear consecutively in the path.
	The lemma follows by Fact~\ref{fct:sp}.\ref{fct5}.
	\qed
\end{proof}

\begin{lemma}\label{lem:sp2}
	Suppose there exists a reconfiguration sequence $\rho$ of length $2m(n+2)$.
	Then $\rho$ visits at least one vertex at every level, and
	all the vertices that it visits at a given level have the same v-state.
\end{lemma}
\begin{proof}
	Since the level of a switched vertex can never increase by more than one,
	and the level of all vertices in $p_b$ is 0 and in $p_e$ is $i + 1$,
	$\rho$ must visit a vertex at every level.

	Let $p$ be the first shortest path in $\rho$ that contains a vertex of level $i$, and let $s$ be the v-state of that vertex.
	Let $p''$ be the first path in $\rho$ after $p$ that does not contain a vertex whose level is $i$ and v-state is $s$,
	and let $p'$ be the path right before that.
	By Lemma~\ref{lem:sppath}, all the paths between $p$ and $p'$ only contain level $i$ vertices whose v-state is $s$.

	Suppose for the sake of contradiction that $p''$ contains a vertex with level $i$ and a v-state that is not $s$,
	and consider the move $(a,b,c,d)$ that created $p''$.
	By Lemma~\ref{lem:sppath} and the fact that $p'$ and $p''$ differ in only one vertex,
	we know that $b$ and $c$ are the only level $i$ vertices in $p'$ and $p''$, respectively.
	Furthermore, Fact~\ref{fct:sp}.\ref{fct5} implies all the edges of $(a,b,c,d)$ must be inter-level.
	We can apply Fact~\ref{fct:sp}.\ref{fct3} to the edges $(b,d), (c,d)$ (if $b$ is odd) or to $(a,b), (a,c)$ (if $b$ is even) to show that
	$cs(b) = cs(c)$.
	However, since $l(c) = l(b)$, Lemma~\ref{lem:sp1} implies $cs(c) = cs(b) + 1$.
	This is a contradiction, and, so, $p''$ does not contain a level $i$ vertex.

	For any path, the set of the levels of its vertices must form a contiguous interval, since an edge can never cross more than one level.
	Because $p''$ has a level $i+1$ vertex (by Lemma~\ref{lem:sp1}) and
	does not have a level $i$ vertex, it therefore only contains vertices with level greater than $i$.
	Lemma~\ref{lem:sp1} further implies that the same holds for all the paths after $p''$ in $\rho$.
	\qed
\end{proof}

Suppose there exists a reconfiguration sequence $\rho$ of length $2m(n+2)$.
Lemma~\ref{lem:sp2} allows us to build an assignment $\theta$
by assigning $\theta_i$ the v-state of the vertices of level $i$ in $\rho$.

\begin{lemma}\label{lem:spsound}
	The assignment $\theta$ is satisfying for $\phi$.
\end{lemma}
\begin{proof}
Consider an arbitrary clause $C_j$, and the vertices at depth $2j -1$.
Each $p\in\rho$ contains exactly one vertex at this depth.
In $p_b$, the c-state of this vertex is 0, while in $p_e$ it is 1, so
there exists some first move $(a,b,c,d)$ with $d(b) = d(c) = 2j-1$ and $cs(b) = 0$ and $cs(c) = 1$.
Since $cs(c) > cs(b)$, Lemma~\ref{lem:sp1} implies that $l(b) = l(c)$.
By Fact~\ref{fct:sp}.\ref{fct4}, either $(b,d)$ or $(c,d)$ is a formula edge, otherwise we would have $0 = cs(b) = cs(d) = cs(c) = 1$.
Furthermore, $(b,d)$ is not a formula edge because $cs(b) = 0$. %and a formula edge always goes from a vertex with c-state 1.
Thus, $(c,d)$ is a formula edge, and we know from the construction that $x_{l(c)} = vs(c)$ satisfies $C_j$.
But $\theta_{l(c)} = vs(c)$ by definition, so $C_j$ is satisfied.
\qed
\end{proof}

We now prove that the reduction is complete.
\begin{lemma}\label{lem:spcomplete}
	If $\phi$ is satisfiable, then there exists a reconfiguration sequence of length at most $2m(n+2)$.
\end{lemma}
\begin{proof}
	\begin{sloppypar}
	Let $\theta \in \bit^n$ be a satisfying truth assignment.
	Let $sat(i,j) = 1$ if $C_j$ is satisfied by $\theta_1, \ldots, \theta_{i}$, and 0 otherwise.
	Let $p(i,k) = s$, $v(i, \theta_i, sat(k,1),1)$, $v(i, \theta_i$, $sat(k,1), 2)$, $\ldots$, $v(i, \theta_i, sat(k,j)$, $2j-1)$, $v(i,\theta_i,sat(k,j), 2j)$,$\ldots,t$.
	We claim that $\rho = p_b, p(1,0), p(1,1)$,$\ldots$, $p(i,i-1), p(i,i)$, $\ldots, p_e$ is a reconfiguration sequence
	of length $2m(n+2)$,
	where the intermediate moves and their length are explained below.
	\end{sloppypar}

	The vertices of $p(i,i)$ can be switched in order of increasing depth using inter-level edges to get $p(i+1,i)$
	in $2m$ steps.
	The paths $p(i,i-1)$ and $p(i,i)$ are different only
	%when there is a difference in value from $sat(i-1,j)$ and $sat(i,j)$ for some $j$.  This can occur only
	when $C_j$ is satisfied by $\theta_i$, in which case there is a formula edge from $v(i,\theta_i, 1, 2j -1)$ to
	$v(i,\theta_i,0,2j)$.
	Using this edge, the vertices of $p(i,i-1)$ can be switched in order of increasing depth to get $p(i,i)$.
	The number of moves required is $2k$, where $k$ is the number of clauses satisfied by $\theta_i$ but not satisfied by $\theta_1,\ldots,\theta_{i-1}$.
	When summed over $\rho$, these add up to at most $2m$,
	since each clause can become satisfied for the first time only once.
	Finally, we can switch between $p_b$ and $p(1,0)$ and between $p(n,n)$ and $p_e$ using in $2m$ steps each.
	\qed
\end{proof}

Combining Lemma~\ref{lem:spsound} and Lemma~\ref{lem:spcomplete} together with the fact that the reduction can
be clearly done in polynomial time, we have the following theorem.
\begin{theorem}\label{thm:hard}
	The \mspr problem is NP-hard, even if $k$ is polynomial in $|V(G)|$.
\end{theorem}

\section{Independent set reconfiguration: the models}\label{sec:rel}
We now turn our attention to reconfiguring independent sets.
Recall that we view an independent set as a set of tokens placed on the vertices such that no two tokens are adjacent.
Consider the reconfiguration rule in which a move from one valid configuration to another is made by a {\it token jump}:
moving a token from one vertex to an unoccupied vertex (not necessarily a neighbor of it), such that the resulting set is independent.
The token sliding, token jumping, and token addition and removal reconfiguration rules give rise to the following
three reconfigurability problems.
\medskip

\noindent\textbf{Token sliding (TS) / token jumping (TJ)}:
Given a graph $G$ and two independent sets $A$, $B$ in $G$, determine if $A$ can be reconfigured into $B$ via a sequence
of independent sets, each of which results from the previous one by a single token slide (for TS) or jump (for TJ).
\smallskip

\noindent\textbf{Token addition and removal (TAR)}: Given a graph $G$, an integer $k$ and two independent sets $A$, $B$ in $G$, both of size $\geq k$, is there a way to transform $A$ into $B$ via independent sets, each of which results from the previous one by adding or removing one vertex of $G$, without ever going through an independent set of size less than $k - 1$?
\medskip

We now establish the equivalence between the TJ and TAR problems. We need some definitions. We say that $A$ and $B$ are TS- (TAR-, TJ-) reconfigurable if they belong to the same connected component of the TS- (TAR-, TJ-) reconfiguration graph. For the three independent set reconfiguration problems, we refer to the corresponding reconfiguration graphs as the TS-graph, TAR-graph, and TJ-graph (of the graph $G$), respectively.
Corresponding reconfiguration sequences will be referred to as TS- (TAR-, TJ-) paths.
Also, we will use the terms token set and independent set interchangeably.

\begin{theorem}\label{thm:TJ-TAR}
Two independent sets $A$ and $B$ of size $s$ in a graph $G$ are TJ-reconfigurable
%belong to the same connected component of the TJ-graph iff they belong to the same connected component of the TAR-graph with parameter $k=s$.
if and only if they are TAR-reconfigurable with parameter $k = s$. Moreover, ${\it dist}_{\rm TAR}(A,B) = 2{\it dist}_{\rm TJ}(A,B)$,
%\todo{define ${\it dist}_{\rm TAR}(A,B)$, $2{\it dist}_{\rm TJ}(A,B)$?}
and there exists an algorithm that, given a reconfiguration sequence between two independent sets in one of these two models outputs a
reconfiguration sequence connecting the two sets in the other model in time polynomial in the length of the sequence. The algorithm maps
every shortest TAR-sequence to a shortest TJ-sequence, and vice versa.
\end{theorem}

\begin{proof}
Any path $P$ in the TJ-graph between two independent sets of size $s$ naturally corresponds to a path $P'$ in the TAR-graph with parameter $k=s$, between the same two sets: Jumping a token from $a$ to $b$ is equivalent to first removing the token from $a$, thereby creating an independent set of size $s-1$, and then adding a token to $b$, resulting in an independent set of size $s$. Applying this transformation to every jump produces, in linear time, a TAR-reconfiguration path $P'$ such that $|P'| = 2|P|$.

Conversely, suppose that $P= (A_0,A_1,\ldots,A_r)$ is a path in the TAR-graph with parameter $k=s$ connecting two independent sets $A=A_0$ and $B=A_r$ of size $s$, where $A\neq B$.  We will  show how to transform $P$ into a  path $P'$ that connects $A$ to $B$ in the TJ-graph. Notice that this is equivalent to finding an $(A,B)$-path in the TAR-graph such that all the token sets are of sizes $k$ or $k-1$.

Let us call an $(A,B)$-reconfiguration path $P= (A_0,A_1,\ldots,A_r)$ in the TAR-graph {\it compressed} if it holds that $A_{i}\neq A_{i+2}$ for all $i = 0,1,\ldots, r-2$. Clearly, every reconfiguration path can be transformed in linear time into a compressed one since if $A_i = A_{i+2}$, then we can remove $A_{i+1}$ and $A_{i+2}$ from the sequence to obtain a shorter TAR-reconfiguration path from $A$ to $B$. We will refer to this transformation as {\it compression}.

%Then, every TAR-reconfiguration path between $A$ and $B$ the reconfiguration path minimizing $\sum_{i}|A_i|$

\begin{sloppypar}
The whole transformation procedure that will produce a TJ-reconfiguration path will consist of a sequence of compressions and {\it peak foldings}, which we define now. Let $P= (A_0,A_1,\ldots,A_r)$ be a compressed $(A,B)$-path in the TAR-graph. For each $i \in \{1,\ldots, r\}$, let $v_i$ denote the token that is either added or removed in the $i$-th step, that is, $A_i\triangle A_{i-1} = \{v_i\}$ (where $\triangle$ denotes the symmetric difference).
Let us call an index $p\in\{1,\ldots, r-1\}$ {\it peak index} if $$|A_p|=|A_{p-1}|+1=|A_{p+1}|+1\ge k+1\,.$$ It follows directly from the definition that for every peak index $p$, we have $A_{p}\backslash A_{p-1}=\{v_p\}$ and $A_{p}\backslash A_{p+1}=\{v_{p+1}\}$.
In particular, $v_p,v_{p+1}\in A_p$, and we can obtain another $(A,B)$-path in the TAR-graph by replacing $A_p$ with the set $A_p':=A_p\backslash \{v_p,v_{p+1}\}$. More formally, {\it peak folding} is the operation of producing from a compressed path $(A_0,A_1,\ldots,A_{p-1},A_p,A_{p+1},\ldots, A_r)$ the path $(A_0,A_1,\ldots,A_{p-1},A_p',A_{p+1},\ldots, A_r)$, where $p$ is a peak index and $A_p'=A_p\backslash \{v_p,v_{p+1}\}$. Since $P$ is compressed, the sets $A_{p-1}$ and $A_{p+1}$ are distinct, which implies that $v_p\neq v_{p+1}$, and in turn \hbox{$v_{p+1}\in A_{p-1}$}. This guarantees that $A_p' = A_{p-1}\backslash \{v_{p+1}\}$ so we can indeed move from $A_{p-1}$ to $A_p'$ with a token removal. Similarly we can see that $A_{p+1} = A_{p'}\cup \{v_{p}\}$.
Notice that since $A_p'$ is a subset of $A_p$, it is independent. Figure~\ref{fig:peak-folding} shows the effect of applying a peak folding on the cardinalities of the token sets.
\end{sloppypar}

\medskip
\begin{figure}[htp]\label{fig:peak-folding}
\begin{center}
\includegraphics[height=27mm]{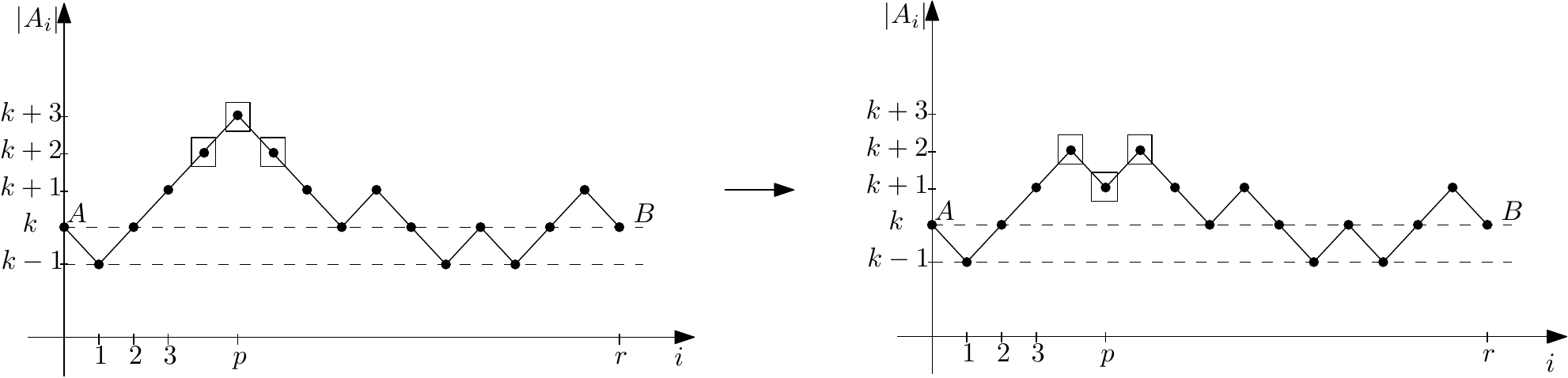}
\caption{Simplifying the reconfiguration path by folding a peak at $p$.}
\end{center}
\end{figure}

Let $P= (A_0,A_1,\ldots,A_r)$ be a given $(A,B)$-path in the TAR-graph. Starting with $P$, we iteratively apply compressions and peak foldings (every peak folding preceded by a compression step), as long as possible, that is, until we obtain a compressed path without peak indices. Notice that every step reduces the total size $\sum_{i}|A_i|$ of the token sets by at least two. All the token set sizes are bounded from below by $k-1$, hence the process eventually stops and produces an $(A,B)$-path $(A_0^*,A_1^*,\ldots, A_m^*)$ in the TAR-graph. Since this path contains no peak indices,
%for every index $i$ such that $|A_i^*|= |A_{i-1}^*|+1 = |A_{i+1}^*|+1$, it holds that $|A_i| =k$, which implies that
it only contains token sets of size $k$ or $k-1$. More specifically, $|A_{i}^*|=k-1$ for odd $i$ and $|A_{i}^*|=k$ for even $i$. In particular, $m$ is even.
%Therefore, for all $i = 0,1,\ldots, m/2-1$, we have  $|A_{2i+2}^*\backslash A_{2i}^*|= 1$ and $|A_{2i}^*\backslash A_{2i+2}^*|= 1$.
The path $(A_0^*,A_2^*,A_4^*,\ldots, A_{m/2}^*)$ is therefore a valid $(A,B)$-path in the TJ-graph. Clearly, the algorithm runs in polynomial time.

Finally, it follows from the above proof that shortest paths get mapped to shortest paths. By the first part of the proof, any $(A,B)$-path of length $r$ in the TJ-graph gets mapped to an $(A,B)$-path of length $2r$ in the TAR-graph. Notice that every shortest $(A,B)$-path in the TAR-graph is compressed, thus, by the second part of the proof, every shortest $(A,B)$-path in the TAR-graph of length $r$ is transformed to an $(A,B)$-path of length $r/2$ in the TJ-graph. This also shows that ${\it dist}_{\rm TAR}(A,B) = 2{\it dist}_{\rm TJ}(A,B)$.
\qed
\end{proof}

Theorem~\ref{thm:TJ-TAR} immediately implies that results holding for the TAR model can be transferred to the TJ model.
New results can also be derived via this relationship:

%\begin{corollary}
%The TJ problem is PSPACE-complete.
%\end{corollary}

\begin{corollary}
There exists a polynomial-time algorithm for the TJ problem in line graphs.
%that given two independent sets in a line graph $G$ determines whether there exists a TJ-reconfiguration sequence between the two sets.
\end{corollary}

\begin{proof}
By Theorem~\ref{thm:TJ-TAR}, the TJ problem in line graphs is polynomially reducible to the TAR problem in line graphs. Due to the correspondence between matchings in a graph and independent sets in its line graph, the problem is polynomially equivalent to the {\sc Matching reconfiguration} problem. For a polynomial-time algorithm for this problem, see Ito et al.~\cite{ItoDHPSUU08}.
\qed
\end{proof}

\section{Hardness of independent set reconfiguration}\label{sec:paths-indep-sets}
TS, TAR, and TJ reconfiguration problems are all PSPACE-complete in general graphs.
For the TS problem, this was announced in~\cite{HearnD05} (see also~\cite{BonsmaC09}) without an explicit proof.
For the TAR problem, this was shown by Ito et al.~\cite{ItoDHPSUU08}.
In fact, their proof uses only token slides (which are done by token additions and removals), also implying that TS is PSPACE-complete.
Theorem~\ref{thm:TJ-TAR} immediately implies PSPACE-completeness for the TJ model.
The hardness of these reconfigurability problems of course implies the hardness of the more difficult problems of finding the length of
the shortest reconfiguration sequence.
However, in this section we show that these related variants remain NP-hard even when the graph is restricted to be perfect.

We use a reduction from the reconfiguration of shortest paths.
Given a graph $G$ and two vertices $s$ and $t$ of $G$ at a distance $k$ apart, let $G_1$ denote the graph obtained from $G$ by deleting from it all the vertices and edges not appearing on any shortest $(s,t)$-path.
For $i\in \{0,1,\ldots, k\}$, let $D_i$ be the set of vertices in $G_1$ at distance $i$ from $s$ and $k-i$ from $t$.
The graph $G'$ is the graph obtained from $G_1$ by turning every set $D_i$ into a clique, and complementing the edges of $G_1$
between every pair of consecutive layers $D_i$ and $D_{i+1}$.
Formally, $V(G') = V(G_1)$ and $E(G') = \{uv:u\neq v,~\exists i$ such that $u,v\in D_i\} \cup \bigcup_{i = 0}^{k-1}\{uv:u\in D_i, v\in D_{i+1}, uv\not\in E(G_1)\}$.
The idea of the construction is that there is a bijective correspondence between shortest $(s,t)$-paths in $G$ and independent sets of size $k+1$ in $G'$.
This gives the following theorem (the proof is straightforward)
 \begin{theorem}
	For every graph $G$, there is a polynomially computable length-preserving bijection (length-doubling for TAR) between shortest reconfiguration sequences in the shortest path reconfiguration graph of $G$
	and those %shortest reconfiguration sequences
	in the TS- (TJ-, TAR-) reconfiguration graph for $G'$.
\end{theorem}
The following corollary is a direct consequence of Theorem~\ref{thm:hard} and the fact that graph $G'$ contains no odd holes or their complements and hence is perfect~\cite{CRST06}. Recall that a {\it hole} in a graph is a chordless cycle with at least four vertices, and a hole is even (odd) if it has an even (odd) number of vertices.

\begin{corollary}
	Let $G$ be a perfect graph, $A,B$ two independent sets in $G$, and $k$ an integer.
	It is NP-hard to determine if there exists a reconfiguration sequence of length at most $k$ between $A$ and $B$ in the TS, TJ, and TAR models,even if $k$ is polynomial in $|V(G)|$.
\end{corollary}

\section{Positive results for independent set reconfiguration}\label{sec:positive}
In this section we identify two restrictions on the input graphs which make the reconfigurability of independents sets easy to solve.

\subsection{Even-hole-free graphs in the TJ model}\label{sec:TJ}
We will show that two token sets of the same size in any even-hole-free graph are TJ-reconfigurable.
Given a graph $G$ and two independent sets $A$ and $B$ in $G$ of the same size, the {\it Piran graph} $\Pi(A,B)$ of $A$ and $B$ is the subgraph of $G$ induced by the vertex set $(A\backslash B)\cup (B\backslash A)$. The following simple lemma gives a sufficient condition under which it is always possible to jump a token from $A$ to~$B$.

\begin{lemma}\label{lemma:lonely-neighbor}
Let $A$ and $B$ be two independent sets of the same size in a graph $G$. If the Piran graph $\Pi(A,B)$ is even-hole-free then there exists a token in $B\setminus A$ with at most one neighbor in $A\setminus B$.
\end{lemma}

\begin{proof}
The Piran graph is bipartite, and as such, it does not contain odd cycles. If in addition,
$\Pi(A,B)$ is even-hole-free, then it must be a forest. Since $|A\setminus B| = |B\setminus A|$, the number of edges in $\Pi(A,B)$ is in fact at most $|A\setminus B|+ |B\setminus A|-1 = 2|B\setminus A|-1$. Therefore there exists a vertex in $B\setminus A$ with at most one neighbor in $A\setminus B$.
\qed
\end{proof}

A consequence of Lemma \ref{lemma:lonely-neighbor} is the following result.

\begin{theorem}\label{thm:even-hole-free}
Let $A$ and $B$ be two independent sets of the same size in a graph $G$.
 If the Piran graph $\Pi(A,B)$ is even-hole-free, then $A$ and $B$ are TJ-reconfigurable.
 Moreover, there exists an algorithm running in time $O(|A|)$ that (if the Piran graph is even-hole-free) finds a shortest TJ-path between the two sets.
\end{theorem}

\begin{proof}
By Lemma~\ref{lemma:lonely-neighbor}, there exists a token in $B\setminus A$ with at most one neighbor in $A\setminus B$. Therefore, the following sequence of token jumps will transform the current independent set $A$ to the target independent set $B$:
\begin{enumerate}
	\item Find a vertex $v$ from $B\backslash A$ with at most one neighbor in the set $A\backslash B$.
	\item If $v$ has a neighbor in $A\backslash B$, say $w$, jump $w$ to $v$.
	\item Otherwise, jump an arbitrary token $w$ from $A\backslash B$ to $v$.
\item Replace $A$ and $B$ with $A\setminus\{w\}$ and $B\setminus \{v\}$, respectively. If $|A|\ge 1$, go to 1.
\end{enumerate}
The whole procedure can be performed in time $O(|A|)$, as follows. We first compute the degrees of the vertices in the ($B\backslash A$)-part of the initial Piran graph. We keep vertices of degree at most one in a queue. Every token jump consists of taking a vertex $v$ from the queue, finding a vertex $w\in A\backslash B$ as specified above, reducing the Piran graph by exactly two vertices ($v$ and $w$), and reducing the degrees of the remaining neighbors of $w$ other than $v$ by one. We add each vertex whose degree drops to one to the queue and repeat the procedure. Since every edge of the initial Piran graph is considered at most once, the running time of the algorithm is $O(|E(\Pi(A,B))|) = O(|A|)$.
Finally, note that since the reconfiguration sequence output by the algorithm consists of $|A\backslash B|$ moves, it is a shortest one.
\qed
\end{proof}

The class of even-hole-free graphs includes the well known class of chordal graphs (hence also trees and interval graphs). The structure of even-hole-free graphs is understood and membership in this class can be decided in polynomial time~\cite{CCKV02}. Notice that if the input graph is even-hole-free, so is the Piran graph. Due to Theorem~\ref{thm:even-hole-free} we can easily solve the TJ reconfiguration problem for the class of even-hole-free graphs. Interestingly, determining the complexity of computing the maximum size of an independent set in an even-hole-free graph is, to the best of our knowledge, an open problem.

The example of the claw $G = K_{1,3}$ with leaves $\{v_1,v_2,v_3\}$, and the independent sets $A = \{v_1,v_2\}$, $B = \{v_1,v_3\}$, shows that the analogue of Theorem~\ref{thm:even-hole-free}
does not hold for the TS model for the whole class of even-hole-free graphs. We leave it as an open question to determine whether the analogue
holds for the class of (claw, even-hole)-free graphs.

\subsection{$P_4$-free graphs in the TS model}\label{sec:TS}

\begin{sloppypar}
In this subsection we give a polynomial time algorithm to solve the TS problem in $P_4$-free graphs.
$P_4$-free graphs (also known as cographs) are graphs without an induced subgraph isomorphic to a 4-vertex path. A polynomial-time algorithm for token sliding in $P_4$-free graphs can be developed based on the following well-known characterization of $P_4$-free graphs~\cite{CLB81}: a graph $G$ is $P_4$-free if and only if for every induced subgraph $F$ of $G$ with at least two vertices, either $F$ or the complement to $F$ is disconnected. A \emph{co-component} of a graph $G=(V,E)$ is the subgraph of $G$ induced by the vertex set of a connected component of the complementary graph \hbox{$\overline G = (V,\{uv~|~u,v\in V,~u\neq v,~uv\not\in E\})$}.
\end{sloppypar}

\begin{theorem}\label{thm:P_4-free}
The TS problem is solvable in time $O(|V|+|E|)$ if the input graph $G=(V,E)$ is $P_4$-free.
Moreover, a shortest reconfiguration sequence, if it exists, can be found in time $O(|V|+|E|)$.
\end{theorem}

\begin{proof}
We claim that Algorithm~1 below solves the TS problem on $P_4$-free graphs.

\begin{algorithm}[ht]%\label{alg:1}
\caption{TS-reconfiguration of independent sets in $P_4$-free graphs}
Input: A $P_4$-free graph $G=(V,E)$ and two independent sets $A$, $B$.\\
Output: A shortest $(A,B)$-path in the TS-graph, if one exists,
%$A$ and $B$ are TS-reconfigurable,
NO otherwise.
\begin{algorithmic}[1]
\STATE {\bf if}~~$|V(G)| = 1$~~{\bf then}~~return the trivial TS-path if $|A| = |B|$, and NO otherwise.
%\ENDIF
\STATE {\bf if}~~$G$ is disconnected, with connected components $C_1,\ldots, C_m$~~{\bf then}
\STATE ~~~~~{\bf if}~~there is an $i\in \{1,\ldots,m\}$ such that $|A\cap C_i|\neq |B\cap C_i|$~~{\bf then}~~return NO.
\STATE ~~~~~{\bf else}~~solve the problem recursively for the connected components $C_1,\ldots, C_m$ \\
~~~~~~~~~~~with respective token sets \hbox{$(A\cap C_1,B\cap C_1), \ldots (A\cap C_m,B\cap C_m)$}. \STATE ~~~~~~~~~~~{\bf if} one of the outputs is NO~~{\bf then}~~return NO.
\STATE ~~~~~~~~~~~{\bf else} merge the corresponding $(A\cap C_i,B\cap C_i)$-paths into an $(A,B)$ \\
~~~~~~~~~~~~~~~~~TS-path $P$, return~$P$.
\STATE {\bf else}~~
\STATE ~~~~~{\bf if}~$|A| = |B|  = 1$~{\bf then}~return an $(A,B)$ TS-path corresponding to a shortest\\
~~~~~~~~~~~$(A,B)$-path in~$G$.
\STATE ~~~~~{\bf else}
\STATE ~~~~~~~~~~~{\bf if} $A$ and $B$ are in the same co-component of $G$~~{\bf then}~~solve the problem \\~~~~~~~~~~~~~~~~for $A$ and $B$ recursively on that co-component and return the output.
\STATE ~~~~~~~~~~~{\bf else} return NO.
\end{algorithmic}
\end{algorithm}

\begin{sloppypar}
The correctness of the algorithm is straightforward, using the above-mentioned characterization of $P_4$-free graphs~\cite{CLB81}. Using the result of Corneil et al.~\cite{CPS85} showing that the decomposition of a $P_4$-free graph $G=(V,E)$ into one-vertex graphs by means of taking components or co-components can be computed in time $O(|V|+|E|)$~\cite{CPS85}, it is also easy to see that the algorithm can be implemented so that it runs in linear time.
\qed
\end{sloppypar}
\end{proof}

Theorem \ref{thm:P_4-free} can be used to prove that the TS problem is solvable in polynomial time if the input graph is (claw, paw)-free (recall that the claw is $K_{1,3}$ and the paw is the graph obtained from the claw by adding one edge).
This is due to the observation that the only connected (claw, paw)-free graph containing an induced $P_4$ are (long enough) paths and cycles.

\section{Concluding remarks}
In this paper, we studied the reconfiguration variant of the shortest path problem. 
We believe that the major open problem is to determine the complexity of deciding whether two shortest paths are reconfigurable. 
If the problem is NP-hard, then it will be the first example of 
a ``natural'' problem in P whose reconfigurability version is NP-hard.
If the problem is polynomially solvable, then it will be the first example of 
an efficiently solvable reconfigurability problem with reconfiguration graphs of large diameter. 

Our results are somewhat orthogonal to previous research on reconfiguration since we consider the length of a shortest bath between two instances in the reconfiguration graph. If we assume that the bound $k$ on the reconfiguration sequence length is given in unary, \mspr is in NP and Theorem~\ref{thm:hard} says it is NP-complete. It would be interesting to analyze whether similar results hold for other problems that have been studied in the context of reconfiguration.\medskip

\noindent {\bf Acknowledgements.}
We are grateful to Takehiro Ito and Daniel Pellicer for interesting and fruitful discussions.
We also thank Paul Bonsma for pointing us to problems in P whose reconfiguration versions are PSPACE-complete.\medskip

\subsection*{Addendum}
Recently, Paul Bonsma proved that shortest path reconfiguration is PSPACE-complete~\cite{PB10}.

\bibliographystyle{plain}
\bibliography{reconf}

\begin{thebibliography}{10}

\bibitem{PB10}
Paul~S. Bonsma.
\newblock Shortest path reconfiguration is {PSPACE}-hard, 2010.
\newblock arXiv:1009.3217v1 [cs.CC].

\bibitem{BonsmaC09}
Paul~S. Bonsma and Luis Cereceda.
\newblock Finding paths between graph colourings: {PSPACE}-completeness and
  superpolynomial distances.
\newblock {\em Theor. Comput. Sci.}, 410(50):5215--5226, 2009.

\bibitem{BonsmaCHJ07}
Paul~S. Bonsma, Luis Cereceda, Jan van~den Heuvel, and Matthew Johnson.
\newblock Finding paths between graph colourings: Computational complexity and
  possible distances.
\newblock {\em Electronic Notes in Discrete Mathematics}, 29:463--469, 2007.

\bibitem{CerecedaHJ08}
Luis Cereceda, Jan van~den Heuvel, and Matthew Johnson.
\newblock Connectedness of the graph of vertex-colourings.
\newblock {\em Discrete Mathematics}, 308(5-6):913--919, 2008.

\bibitem{CerecedaHJ09}
Luis Cereceda, Jan van~den Heuvel, and Matthew Johnson.
\newblock Mixing 3-colourings in bipartite graphs.
\newblock {\em European Journal of Combinatorics}, 30:1593--1606, 2009.

\bibitem{CRST06}
Maria Chudnovsky, Neil Robertson, Paul Seymour, and Robin Thomas.
\newblock The strong perfect graph theorem.
\newblock {\em Ann. of Math.}, 164:51--229, 2006.

\bibitem{CCKV02}
Michele Conforti, G\'erard Cornu\'ejols, Ajtai Kapoor, and Kristina
  Vu\v{s}kovi\'c.
\newblock Even-hole-free graphs part {II}: Recognition algorithm.
\newblock {\em J.~Graph Theory}, 40:238--266, 2002.

\bibitem{CLB81}
Derek~G. Corneil, H.~Lerchs, and L.~Stewart Burlingham.
\newblock Complement reducible graphs.
\newblock {\em Discrete Applied Mathematics}, 3(3):163--174, 1981.

\bibitem{CPS85}
Derek~G. Corneil, Yehoshua Perl, and Lorna~K. Stewart.
\newblock A linear recognition algorithm for cographs.
\newblock {\em SIAM J.~Comput.}, 14(4):926--934, 1985.

\bibitem{GopalanKMP09}
Parikshit Gopalan, Phokion~G. Kolaitis, Elitza~N. Maneva, and Christos~H.
  Papadimitriou.
\newblock The connectivity of {B}oolean satisfiability: Computational and
  structural dichotomies.
\newblock {\em SIAM J. Comput.}, 38(6):2330--2355, 2009.

\bibitem{HearnD05}
Robert~A. Hearn and Erik~D. Demaine.
\newblock {PSPACE}-completeness of sliding-block puzzles and other problems
  through the nondeterministic constraint logic model of computation.
\newblock {\em Theor. Comput. Sci.}, 343(1-2):72--96, 2005.

\bibitem{ItoDHPSUU08}
Takehiro Ito, Erik~D. Demaine, Nicholas J.~A. Harvey, Christos~H.
  Papadimitriou, Martha Sideri, Ryuhei Uehara, and Yushi Uno.
\newblock On the complexity of reconfiguration problems.
\newblock In {\em ISAAC}, volume 5369 of {\em Lecture Notes in Computer
  Science}, pages 28--39. Springer, 2008.

\bibitem{ItoKD09}
Takehiro Ito, Marcin Kami\'nski, and Erik~D. Demaine.
\newblock Reconfiguration of list edge-colorings in a graph.
\newblock In {\em WADS}, volume 5664 of {\em Lecture Notes in Computer
  Science}, pages 375--386. Springer, 2009.

\bibitem{IWOCA2010}
Marcin Kami\'nski, Paul Medvedev, and Martin~Milani\v c.
\newblock Shortest paths between shortest paths and independent sets.
\newblock In {\em IWOCA, {\it Lecture Notes in Computer Science}}, volume TBD
  of {\em Lecture Notes in Computer Science}. Springer, 2010.

\end{thebibliography}
\end{document}